\documentclass[10pt,english,journal]{IEEEtran}

\usepackage[T1]{fontenc}
\usepackage[utf8]{inputenc}
\usepackage{geometry}
\usepackage{amsmath}
\usepackage{xpatch}
\usepackage{amssymb}
\usepackage{esint}
\usepackage{mathtools}
\usepackage{amsthm}
\usepackage{nicefrac}
\usepackage{bigints}
\usepackage{mathtools}
\usepackage{blkarray, bigstrut}
\usepackage{physics}
\usepackage{calligra}
\usepackage{graphicx}
\usepackage{epstopdf}
\usepackage{dsfont}
\usepackage{array,ragged2e}
\usepackage{enumitem}
\usepackage{etoolbox}
\usepackage{babel}
\usepackage[nopar]{lipsum}
\usepackage{psfrag}
\usepackage[ruled,lined,linesnumbered]{algorithm2e}
\usepackage{algorithmic}
\usepackage{algorithm2e}
\usepackage{balance}
\usepackage{float}
\usepackage{subcaption}

\SetKw{KwBy}{by}

\makeatletter
\newcommand{\removelatexerror}{\let\@latex@error\@gobble}
\makeatother

\graphicspath{ {Figures/} }

\xpatchcmd{\proof}{\hskip\labelsep}{\hskip5\labelsep}{}{}  
\makeatletter
\xpatchcmd{\proof}{\@addpunct{.}}{\@addpunct{:}}{}{}
\makeatother

\renewcommand\[{\begin{equation}}
\renewcommand\]{\end{equation}} 
\usepackage{listings}
\usepackage{fancyvrb}
\usepackage{framed}

\usepackage{courier}
\usepackage[usenames,dvipsnames,table]{xcolor}

\definecolor{dkgreen}{rgb}{0,0.3,0}
\definecolor{gray}{rgb}{0.5,0.5,0.5}

\makeatletter
\newcommand*{\rom}[1]{\expandafter\@slowromancap\romannumeral #1@}
\makeatother

\usepackage{siunitx}
\usepackage{tabu}
\usepackage{booktabs}
\usepackage{multirow}
\usepackage{capt-of}
\usepackage{array}
\usepackage{arydshln}
\newcommand{\comment}[1]{}
\begin{document}

\title{
A Federated DRL Approach for Smart Micro-Grid Energy Control with Distributed Energy Resources
}

\author{Farhad Rezazadeh$^{1,2}$, Nikolaos Bartzoudis$^1$\\
{\normalsize{} $^1$ Telecommunications Technological Center of Catalonia (CTTC), Barcelona, Spain\\ $^2$ Technical University of Catalonia (UPC), Barcelona, Spain}\\  
{\normalsize{}Contact Emails: \texttt{\{frezazadeh, nbartzoudis\}@cttc.es}}}

\maketitle

\begin{abstract} 
The prevalence of the Internet of things (IoT) and smart meters devices in smart grids is providing key support for measuring and analyzing the power consumption patterns. This approach enables end-user to play the role of prosumers in the market and subsequently contributes to diminish the carbon footprint and the burden on utility grids. The coordination of trading surpluses of energy that is generated by house renewable energy resources (RERs) and the supply of shortages by external networks (main grid) is a necessity. This paper proposes a hierarchical architecture to manage energy in multiple smart buildings leveraging federated deep reinforcement learning (FDRL) with dynamic load in a distributed manner. Within the context of the developed FDRL-based framework, each agent that is hosted in local building energy management systems (BEMS) trains a local deep reinforcement learning (DRL) model and shares its experience in the form of model hyperparameters to the federation layer in the energy management system (EMS). Simulation studies are conducted using one EMS and up to twenty smart houses that are equipped with photovoltaic (PV) systems and batteries. This iterative training approach enables the proposed discretized soft actor-critic (SAC) agents to aggregate the collected knowledge to expedite the overall learning procedure and reduce costs and CO2 emissions, while the federation approach can mitigate privacy breaches. The numerical results confirm the performance of the proposed framework under different daytime periods, loads, and temperatures.

\end{abstract}

\section{Introduction}
\IEEEPARstart{S}{mart} grid assisted by distributed renewable energy generation, IoT, and AI solutions is an emerging trend to ameliorate the reliability and efficiency of the power systems while impending environmental pollution and energy crisis. The digitization of smart grids pursuing AI-driven solutions and using local renewable energy can enable users to control their energy consumption and participate as prosumers in the energy market ecosystem. The old energy systems benefit from large centralized power generation with unidirectional mechanisms, whereas the smart energy solution leverages distributed local clean energy generation where supply and demand can be handled in a dynamic and shared environment.

The local generation in the form of solar PV systems that convert solar energy into electrical energy is dependent on the presence of solar radiation, weather, and temperature conditions. This kind of renewable energy system has high availability and easy installation. Nevertheless, the integration of the renewable energy system to the main grid can be challenging for aggregators and controllers due to their intermittent behaviors and instability. In this intent, the AI solutions can pave the way to efficiently optimize the flow of energy while maximizing the utilization at peaks of consumption with accurate forecasting, effective storage, and trading energy. The main contributions of our paper can be summarized as follows:
\begin{itemize}
    \item The energy control problem is treated as an optimization problem, focusing on minimizing the external energy supply and subsequently decreasing the cost and CO2 emissions where AI agents learn to take appropriate actions to charge or discharge batteries and trade energy in different conditions. 
    \item The centralized energy and load control can lead to privacy breaches where users should provide information on consumption and generated energy. To address this, we proposed an FDRL approach to develop a decentralized DRL \cite{ans-far} algorithm based on the actor-critic method \cite{cont1}. This approach enables the households to share their experiences in the form of local neural network parameters to the federation layer and mitigate the households' privacy concerns.
    \item The implemented distributed learners leverage a combination of deep Q-network (DQN) \cite{cont2} and policy gradients \cite{cont3} in the form of stochastic actor-critic \cite{cont4}. We propose a discretized version of SAC \cite{cont5} to surmount the curse of dimensionality concerning the inordinate large state space of the defined problem and the presence of multidimensional user load information. This approach also reduces the need for hyperparameter tuning and stabilization of the learning procedure \cite{cont6}. Each local SAC agent in BEMS derives the optimal policy energy control under different daytime periods, loads, and temperatures.
    \item We have validated our openAI Gym-based \cite{cont7, scal_ac} hierarchical architecture and assess its capabilities in realistic scenarios through an exhaustive simulation campaign, accounting for up to 20 smart houses.

\end{itemize}

The remainder of this paper is organized as follows:
Sec.~\ref{sec:related} provides an overview of the related works in the field.
Sec.~\ref{sec:Preliminaries} highlights the preliminaries of DRL algorithm.
Sec.~\ref{sec:scenario}  elaborates the framework overview considered scenario, the proposed SAC algorithm, and describing the interaction among agent and environment.
Sec.~\ref{sec:perf_eval} explains the performance evaluations of proposed approach.
Finally, Sec.~\ref{sec:conclusion} provides the final remarks and concludes this paper.
\section{Related Work}
\label{sec:related}
AI-driven approaches applied to the smart grid have recently gained momentum in distributed energy resource control tasks. In this context, federated learning (FL) stands out among a multitude of different approaches and is at the center of strong research interest. The authors of \cite{FL1} propose a novel FL solution assisted with AI of Things (AIoT) and edge-cloud collaboration for efficient personal energy data sharing in smart grids. Moreover, they devise a DRL-based incentive algorithm with two layers in the presence of multidimensional user private information and the large state space to optimal training strategies of energy data owners (EDOs). The simulations have demonstrated that the proposed scheme can efficiently motivate EDOs' high-quality local model sharing while reducing task latencies. In \cite{FL2}, the authors propose a demand response algorithm for residential users, while accounting for uncertainties in the load demand and electricity price, users' privacy concerns, and power flow constraints imposed by the distribution network. To address the uncertainty issues, they develop a DRL algorithm using an actor-critic method and apply FL. They show that the proposed load scheduling algorithm can benefit the load aggregator by a 33\% reduction in the aggregate demand during peak hours. It also provides benefits for users by reducing 13\% the expected daily cost. In \cite{FL3}, the authors propose a systematic FL framework, that addresses both the horizontal and vertical separation of data in a distributed fashion. To protect user privacy and secure power traces, raw data are always stored locally to prevent data leakage, while the model parameters and sample statistics are Paillier-encrypted before the exchange. The authors of \cite{FL4} 
propose a system model using edge computing and FL to tackle data diversity challenges related to short-term load forecasting in the smart grid. This decentralized scheme allows for an increase in the volume and diversity of data used to train deep learning models. The simulations have demonstrated a promising approach to
create highly performing models with a significantly reduced
networking load compared to a centralized model. Also \cite{FL5} propose a distributed approach that
leverages FDRL to manage the optimal energy consumption of
multiple smart houses. The simulation results have shown that the proposed house BEMS framework successfully managed the energy consumption of multiple smart houses within a moderate number of training iterations while providing faster convergence compared to the non-FDRL approach.

\section{Preliminaries}
\label{sec:Preliminaries}
The abbreviations used in this paper are summarized in
Table~\ref{tab:notations}.

\begin{table}[!t]
\caption{Abbreviations}
\label{tab:notations}
\centering
\begin{tabular}{@{}lc@{}}\toprule
\textbf{Abbreviation} & \textbf{Description} \\ \midrule
$IoT$ & Internet of things\\ \hdashline
$RERs$ & Renewable energy resources\\ \hdashline
$FDRL$ & Federated deep reinforcement learning \\ \hdashline
$BEMS$ & Building energy management systems\\ \hdashline
$DRL$ & Deep reinforcement learning\\ \hdashline
$EMS$ & Energy management system\\ \hdashline
$PV$ & Photovoltaic\\ \hdashline
$SAC$ & Soft actor-critic\\ \hdashline
$DQN$ & Deep Q-network\\ \hdashline
$FL$ & Federated learning \\ \hdashline
$MDP$ & Markov decision process\\ \hdashline
$TD$ & Temporal difference\\ \hdashline
$DNN$ & Deep neural network\\ \hdashline
$ML$ & Machine learning\\ \hdashline
$DDPG$ & Deep deterministic policy gradient\\
\bottomrule
\end{tabular}
\end{table}
\subsection{Reinforcement learning}

The mathematical context of Markov decision process (MDP) can be considered as a formal framework to formalize diverse RL methods for learning optimum decision-making policies in a fully or partially observable environment. One of the important issues is the intelligent agent can dynamically make decisions based on actions, states, and rewards where the states refer to possible network configurations and reward (or penalty) stands for feedback signal from the network (environment) that implies the agent's performance. The agent observes the state of the network at each time step $t$ and acts on that network to transit from one state to another. Typically, an MDP is defined by a 5-tuple $(S, A, P, \gamma, R)$ where $S$ is a set of state (state space), $A$ refers to a set of action (action space), $P$ denotes the transition probability from current state $s$ to the next state $s^\prime$ where they govern rules of state transitions and define our dynamics. The $R$ notation stands for the reward function. The agent obtains an immediate reward  by taking action $a_t$ in state $s_t$. Unlike the finite or episodic tasks, in continuing tasks the total reward can be infinite. The $\gamma$ defines reward discounting hyperparameter that results in significant deviations in the performance of agent where the future agent values rewards will turn to less and less. Indeed, it is a real-valued discount factor weighting to determine the importance of future rewards as a short-sighted or myopic agent $(\gamma=0)$, i.e., the aim of agent is to maximize its current/immediate rewards, or far-sighted agent $(\gamma=1)$ that strives to accumulate long-term higher rewards. The $\gamma$ is often chosen in $[0.95, 0.99]$. The total discounted rewards from time step $t$ is given by
\begin{equation}
{G}_t = R_{t+1}+\gamma R_{t+2}+\gamma^2 R_{t+3}+ ...  \sum_{n=0}^{\infty}\left(\gamma^{n} R_{t+n+1}\right).
\end{equation}
We can derive useful recursive relation between rewards and subsequent time step, $G_t = R_{t+1} + \gamma G_{t+1}$ that it can attenuate computational complexity and memory requirements. However, there is one more piece we need to complete the MDP. The key term of MDP is decision. Indeed, the way we make decisions for what actions to do in what states is called a policy which denotes with the symbol $\pi$ that it is usually a probabilistic function as mapping of states to actions.


The excellence of each state or state-action pair can be determined by how large a future reward of the agent. The state value function for the policy $\pi$ is an explicit measure of how good the state or how much reward to expect
\begin{equation}
{V}_{\pi}(s)= \mathbb{E}_{\pi}\left[\sum_{n=0}^{\infty}\left(\gamma^{n} R_{t+n+1} | S_t = s \right)\right],
\end{equation}
and expectation value of the reward at time $t$ is defined as action value function
\begin{equation}
{Q}_{\pi}(s,a) = \mathbb{E}_{\pi}\left[\sum_{n=0}^{\infty}\left(\gamma^{n} R_{t+n+1} | S_t = s, A_t = a \right)\right].
\end{equation}
According to total expectation in reverse and nested expected value, we can define Bellman equation by the assumption of MDP
\begin{equation}
{V}_{\pi}(s)= \sum_{a}\pi(a|s)\sum_{s^\prime}\sum_{r}p(s^\prime, r | s,a)[r + \gamma{V}_{\pi}(s^\prime)].
\end{equation}
The above recursive relationship between each value function and the value function of successor state is considerd as heart of RL to solve stochastic and non-deterministic search problem where agent encounter random events. Notice in deterministic policy, there is just one action with $\pi(a|s) = 1$ and the rest receive $0$ value. There are some algorithms that specifically apply to only the state-value function or only the action-value function. Let suppose we have two policies $\pi_1$ and $\pi_2$ and $\pi_1$ is better than $\pi_2$, if expected return of $\pi_1$, $V_{\pi_1}(s)$ greater than or equal to the expected return of $\pi_2$, $V_{\pi_2}(s)$ for all states
\begin{equation}
\forall\quad s \in S,\quad if\quad V_{\pi_1}(s) \geq V_{\pi_2}(s)\quad \rightarrow \quad\pi_1 \geq \pi_2
\end{equation}

The optimal policy in RL is the best policy for which there is no greater value function.
In the parlance of RL, the agent engages in exploration and exploitation where exploitation or greed refers to take the best-known action while taking sub-optimal action is called exploration (i.e. the way of exploring the environment). This process of exploration and exploitation enables the agent to refine its model and gradually find the true estimate of the value function.

Q-learning is a model-free approach in the class of temporal difference (TD) learning algorithms where it has an online nature and updates the agents' estimate of value function at each time step. To update the value function concerning TD we have
\begin{equation}
{V}(s_t)= {V}(s_t) + \alpha [R_{t+1}+\gamma V(s_{t+1}) - V(s_t)]
\end{equation}
where  $\alpha$ denotes exponentially smooth mean approach to deal with non-stationary distributions instead of calculating sample mean inefficiently and $R_{t+1}+\gamma V(s_{t+1})$ is target of TD method and $V(S_t)$ refers to old estimate. Moreover, the quantity in the brackets $R_{t+1}+\gamma V(s_{t+1}) - V(S_t)$ is called TD error and therefore we have squared error $E = [Target-Prediction]^2$ with pursuing a gradient descent approach and the aim is the prediction value be closer to target value. This approach that use one estimate to update another estimate is called bootstrapping. The updated Q-value (quantifiable value for more lucrative action) in Q-learning as off-policy method is given by
\begin{multline}
Q(s_t,a_t)=
Q(s_t,a_t) +\alpha [R_{t+1}+\gamma \underset{a}{\max}Q(s_{t+1}, a_{max})\\
- Q(s_t,a_t)].
\end{multline}

\subsection{Deep Reinforcement Learning}
The Q-learning works in a very simplistic environment while in practice, communication network problems have complicated system models with large and continuous state spaces that can have an infinite number of states. Therefore, the tabular representation of the action-value function in the previous section leads to computationally complexity and also training time limitation problems. Unlike the tabular and non-parametric approach, RL is assisted with deep neural network (DNN) in DRL to surmount the curse of dimensionality with respect to inordinate large state spaces. Indeed, It is possible to model $Q(s,a)$ based on simple linear regression but another new approach is the DNN that benefits from crafting inductive biases to overcome effectively the curse of dimensionality and thereby enables RL to scale the decision-making problems for intractable high-dimensional state and action spaces through approximate the Q-value function. The DQN takes the state as input and returns approximated Q-functions of all actions under the input state. In this regard, finding a convincing solution for the instability of function approximation techniques in RL-based agents is a necessity. In his regard, we parameterize the value function with linear regression or a neural network. Let define $V(s) = \theta^T s$ where $s$ is a feature vector which represents the state and in linear regression for calculating the value function prediction $\theta$ refers to the weights. According to the previous section, we apply gradient descent using squared error between target and prediction. Instead of updating $V(s)$ directly, the goal is to update the parameters of $V$ that according to the chain rule and multiply to the gradient of $V(s)$ we have 
\begin{equation}
\theta \leftarrow \theta + \alpha [r+\gamma V(s^\prime) - V(s)]\frac{\partial V(s)}{\partial\theta}
\end{equation}
and in linear regression, we have a simple update rule 
\begin{equation}
\theta \leftarrow \theta + \alpha [r+\gamma V(s^\prime) - V(s)]s
\end{equation}
with the same logic for $Q-updating$, we just consider $s$ as input and for infinite action, the number of output is equal to the number of possible actions
\begin{equation}
\theta \leftarrow \theta + \alpha [r+\gamma \underset{a}{\max}Q(s^\prime, a_{max}) - Q(s,a)]\frac{\partial Q(s,a)}{\partial\theta}.
\end{equation}

\begin{figure*}[h]
\centering
\includegraphics[clip, trim = 0cm 0cm 0cm 0cm,width=12cm]{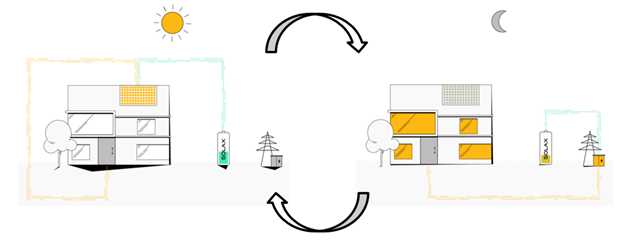}
\caption{\small The proposed single house system in different daytime and temperature.}
\label{fig:sen_PROBLEM}
\end{figure*}

\section{Proposed FDRL Framework}
\label{sec:scenario}

\subsection{System Description}
\begin{figure}[h]
\centering
\includegraphics[clip, trim = 0cm 0cm 0cm 0cm,width=8cm]{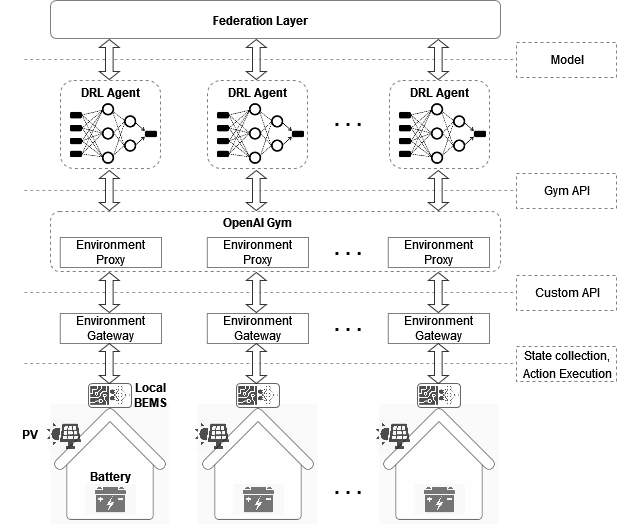}
\caption{\small The proposed hierarchically distributed framework.}
\label{fig:FDRL}
\end{figure}

We consider the proposed framework as combination  of FL and DRL techniques in the form of FDRL. FL is a decentralized machine learning (ML) methodology that increases the amount and variety of data used to train DNN models to help the micro-grid EMS to predict load and control energy consumption optimally. Fig.~\ref{fig:sen_PROBLEM} shows the proposed system (local nodes/house) includes battery settings, pre-processing of power consumption, and production data where the energy is supplied by a local solar panel (considering different air temperatures) or a battery and also by an external network (to cover deficiencies). As depicted in Fig.~\ref{fig:FDRL}, within the context of our FDRL-based framework each agent trains a local DRL model and shares its experience, under the form of model hyperparameters, to those entities belonging to the corresponding federation layer. This iterative training approach enables federation layer to aggregate the collected knowledge of single agents into a global updated model, stored into an EMS to allow faster feedbacks. Within the smart grid domain, we envision the dynamic setup of a network of local BEMS able to access the local house agent information and extract local knowledge without the need for a centralized entity performing decisions on the aggregated information. 

Let us introduce a micro-grid composed of a single fedeartion layer (hosted in EMS) and a of set $\mathcal{H}$ of houses, wherein a set of BEMS $\mathcal{L}$ is deployed. Each house $h\in\mathcal{H}$ is equipped with a PV system and a battery $\mathcal{B}$ with capacity $h_c$. Let us consider a time-slotted system where time is divided into decision intervals $t\in\mathcal{T}=\{1,2,\ldots,T\}$. The action can be taken only at the beginning of each decision interval. In order to enhance efficiency and avoid communication overheads, we allow the federation layer to collect the local models (and share the updated ones) only every $\hat{T}$ decision intervals, defining this time period as the federation episode. Different strategies can be adopted to derive the global federated model, each one implementing a predefined federation strategy function that we use an average model. Let consider $\omega_{h}$ weight of local DRL agent in EBMS and then $\omega_{G}$ as aggregated and global model of $\mathcal{H}$ EBMS in the form of $\omega_G =\frac{1}{\mathcal{H}} \sum_{h=1}^{\mathcal{H}}\omega_h$. The benefit coming from our approach is three-fold: i) it enables power control at the edge of the micro-grid environment, thus accounting for more timely and accurate information, ii) the amount of control information that needs to cross the micro-grid to reach the central controller is dramatically decreased, thus reducing the overhead towards the EMS and avoiding bottlenecks, iii) by allowing information exchange among local BEMS, we enable the provisioning of FL schemes to further enrich the capabilities of the DRL agents by improving the generalization of the learning procedure.  

All together, we can formalize our problem as: $min\ \sum_{t=1}^{T}\ E\left[\sum_{h\in H} e_h^{\left(t\right)}\right]$ where $e_h^{\left(t\right)}$ refers to the amount of energy that is supplied from external network with respect to PV system and battery.  We formulate the optimization problem as an MDP. Our objective is to minimize the external energy supply and subsequently decrease the cost and CO2 emissions under predefined threshold and constraints, where AI agents learn to take appropriate actions to charge or discharge batteries and trade energy in different conditions. In order to define the reward setting of MDP as close as possible to our models, we propose a reward-penalty technique in implementation.For the \emph{action space}, the local DRL agents (i.e., BEMS) consider three main actions: i) Trading the surpluses and shortages of electricity with an external network where they ignore the role of battery ($a_1$), ii) Charging battery by the surplus of electricity production ($a_2$), and iii) Discharging the battery when consumption is greater than production to compensate required energy ($a_3$) and then we have $\mathcal{A} = \{a_1, a_2, a_3\}$. Indeed, the proposed FDRL framework targets the optimization of energy costs under a long-term view by choosing the best action in different states of micro-grid environments. The \emph{state space} includes the following conditions i) generated energy by PV system ($s_1$), ii) current battery capacity ($s_2$), iii) temperature based on daytime ($s_3$), and iv) amount of power consumption ($s_4$). We can define \emph{state space} as $s_i^{(t)} = \{ ( a_{1,h}^{(t)}, a_{2,h}^{(t)}, a_{3,h}^{(t)}, a_{4,h}^{(t)} ) \mid \forall h \in \mathcal{H} \}$. As we mentioned in this section,  the main idea for the reward function is to maximize the difference of generated or stored energy with the supplied energy by the external network that can be defined as $r^{(t)}_h \in \mathcal{R}$ for $h$-th agent as $r^{(t)}_h = \frac{\nu_h^{(t)}}{e_h^{(t)}}$, where $\nu_h^{(t)}$ refers to amount of energy that is generated by PV or stored in battery and supply energy inside house. 

\subsection{Local Training Algorithm}
SAC is an actor-critic method and combination of policy optimization and Q-Learning. Let define $\rho_\pi(s_t)$ and $\rho_\pi(s_t, a_t)$ to denote the state and state-action distribution respectively that induced by policy $\pi$ in micro-grid environment $\pi(a_t|s_t)$. Unlike the DDPG \cite{DDPG} approaches, SAC algorithm is based on stochastic policy gradient \cite{cont4}. The basic idea behind policy-based algorithms is to adjust the parameters $\phi$ of the policy in the direction of the performance gradient $\nabla_{\phi}J(\pi_{\phi})$ concerning the policy gradient theorem \cite{cont3}. The objective for finite-horizon MDPs is given by, $J_{\pi} = \mathbb{E} \left[\sum_{i=t}^{T}\gamma^{i-t}[r_i+\alpha\hat{\mathcal{H}}(\pi(\cdot|s_i))]  \right]$. As we mentioned before, $\gamma$ is the discount factor. The temperature parameter $\alpha$ determines the relative importance of the $\hat{\mathcal{H}}$ as policy entropy term against the reward, thereby handle the stochasticity of the optimal policy. Let us define entropy-augmented accumulated return or soft return as $G_t = \sum_{i=t}^{T}\gamma^{i-t}[r_i-\alpha\log\pi(a_i|s_i)]$. Then we can define soft Q-value with respect to policy $\pi$ as $Q_{\pi} (s_t, a_t) = \mathbb{E}[r]+\gamma\mathbb{E}[G_{t+1}]$.
To stabilize the learning \cite{TD3} and mitigate meticulous hyperparameter tuning we use the (clipped) double Q-learning technique \cite{mano-far} to parameterizes critic networks and critic targets by ${\theta_1}$, ${\theta}_2$ and ${\theta}_1^{\prime}$,${\theta}_2^{\prime}$ respectively. In our method, we store ${({s}_t,{a}_t,{r}_t,{s}_{t+1})}$ to train deep Q-Network and sample random many batches from the experience replay $\beta$ (buffer/queue) as training data. We take a random batch $B$ for all transitions ${({s}_{t_{B}},{a}_{t_{B}},{r}_{t_{B}},{s}_{t_{B}+1})}$.

\begin{algorithm}[t]
\caption{Local training algorithm}
\scriptsize
\SetAlgoLined
 Initialize actor network $\phi$ and critic networks $\theta_1$, $\theta_2$
 
 Initialize (copy parameters) target networks ${\theta}_1^{\prime}$, ${\theta}_2^{\prime}$

 Initialize learning rate $\ell_{\alpha}, \ell_Q, \ell_{\pi}$
 
 Initialize replay buffer $\beta$
 
 Import custom micro-grid environment (`CTTC--v2')
 
 done = False
 
 t=0
 
 \While {t < T}{

  \eIf{t < start\_timesteps}{
   $a$ = env.action\_space.sample()
   }{
   Select action $a\sim\pi_{\phi}(a|s)$
   
  }
  next\_state, reward, done, \_ = env.step($a$)
  
  store the new transition ${({s}_t,{a}_t,{r}_t,{s}_{t+1})}$ into $\beta$
  
   \If{t $\geq$ start\_timesteps}{
   sample batch of transitions ${({s}_{t_{B}},{a}_{t_{B}},{r}_{t_{B}},{s}_{t_{B}+1})}$
   
    $\theta_i\longleftarrow\theta_i-\ell_{Q}\nabla_{\theta_i}J_Q(\theta_i)$, \quad i=1,2 \quad\#Update soft Q-function

   \If{$mod(t,freq)$ == 0}{ 
   $\phi\longleftarrow\phi+\ell_{\pi}\nabla_{\phi}J_{\pi}(\phi)$ \quad \#Update policy weights
   
   $\alpha\longleftarrow\alpha-\ell_{\alpha}\nabla_{\alpha}J(\alpha)$\quad \#Adjust temperature
   
${\theta}_{i}^{\prime}\longleftarrow\tau{\theta}_{i}+(1-\tau){\theta}_{i}^{\prime}$\quad i=1,2\quad \#Update target network
                    }
   }
  \If{done}{
   obs, done = env.reset(), False
   
   \If{$mod(t,\hat{T})$==0}{
   Upload $\omega_{h}^{(t)}$ \;
    Wait for Algorithm~\ref{algo:FDRL}\;
    \emph{ \#Get FL model and update the local one}\;  
    $\omega_{h}^{(t+1)} \leftarrow \omega_{G}^{(t)}$\;
   
   }
   }
  t=t+1
 }
\label{algo:local}
\end{algorithm}

\begin{algorithm}[t]
\scriptsize
\SetKwInOut{Input}{Input}
\SetKwInOut{Output}{Output}
\SetKwInOut{Return}{return}
\Input{$t, T, \omega_{h}^{(t)}$ $\forall h \in \mathcal{H}$}
\Output{Improved federation models $\omega_{G}^{(t+1)}$}
 
  \If{$mod(t,\hat{T})==0 \land t > 0$}{
          \For{ each $\omega_{h}^{(t)}$ $\forall h \in \mathcal{H}$, in parallel}{
                Collect $\omega_{h}^{(t)}$\;
                $\omega_{G}^{(t+1)} \leftarrow \frac{1}{\mathcal{H}} \sum_{h=1}^{\mathcal{H}}\omega_h$\;
                $\omega_{h}^{(t+1)} \leftarrow \omega_{G}^{(t+1)}$\;
         }
        \emph{ \#Return updated local models}\;  \
        \Return{$\omega_{h}^{(t+1)}, \forall h \in \mathcal{H}$}
    }
    Run Algorithm 1\;

\caption{Weight update process in federation layer}
\label{algo:FDRL}
\end{algorithm}

Let us define $Q_{\theta}(s,a)$ and $\pi_{\phi}(a|s)$ as parameterized functions to approximate the soft Q-value and policy, respectively. We consider a pair of soft Q-value functions $(Q_{\theta_1}, Q_{\theta_2})$ and separate target soft Q-value functions $(Q_{\theta^{\prime}_1}, Q_{\theta^{\prime}_2})$. We calculate the update targets of $Q_{\theta_1}$,  $Q_{\theta_2}$ according to $
y= r+\gamma(\underset{i = 1,2}{\min}Q_{\theta^{\prime}_i}(s^{\prime}, a^{\prime})) - \alpha\log\pi_{\phi}(a^{\prime}|s^{\prime}), \quad a^{\prime}\sim\pi_{\phi}
$
we can train soft Q-value by directly minimizing  \cite{mano-far, 6g-far},
\begin{equation}
    J_Q(\theta_i) =\mathbb{E} [(y-Q_{\theta_i}(s,a))^2], \quad i=1,2
\end{equation}
Therefore, the policy update gradients with respect to experience replay ($\beta$) is given by \cite{cont4},
\begin{equation}
\begin{aligned}
\nabla_{\phi}J_{\pi}(\phi) =
\mathbb{E} [-\nabla_{\phi}\alpha\log(\pi_{\phi}(a|s))+(\nabla_aQ_{\theta}(s,a)\\-\alpha\nabla_a\log(\pi_{\phi}(a|s))\nabla_{\phi}f_{\phi}(\xi;s))]
\end{aligned}
\end{equation}

We can update temperature $\alpha$ by minimizing the following objective $J(\alpha)=\mathbb{E}[-\alpha\log\pi_{\phi}(a|s)-\alpha\hat{\mathcal{H}}]$. We follow \cite{dIsac} to derive a discrete action version of the above SAC algorithm. The proposed approach is summarized in Algorithm 1 and ~\ref{algo:FDRL}.

\section{Performance Evaluation}
\label{sec:perf_eval}

In this section, we evaluate our proposed architecture numerical simulations on a dedicated server, equipped with two Intel(R) Xeon(R) Gold 5218 CPUs @ 2.30GHz and two NVIDIA GeForce RTX 2080 Ti GPUs. Moreover, the DNNs are implemented on TensorFlow-GPU version 2.5.0. For the sake of simulating a more realastic scenario, each house uses a mini-dataset from a collected big dataset\footnote{https://github.com/antoine-delaunay/DRL\_SmartGrid/tree/main/Data} correspond to measurement real data of a house between 15/07/2016 and 15/07/2019. It includes, as input features (state values), the values of air temperature, total electricity consumption, and generated energy by PV system per five minutes. First, we compare the performances of different DRL algorithms and random approach when dealing with different conditions such as charging, discharging, and trading, without involving FL. 
\begin{figure}[h]
\centering
\includegraphics[clip, trim = 0cm 0cm 0cm 0cm,width=8cm]{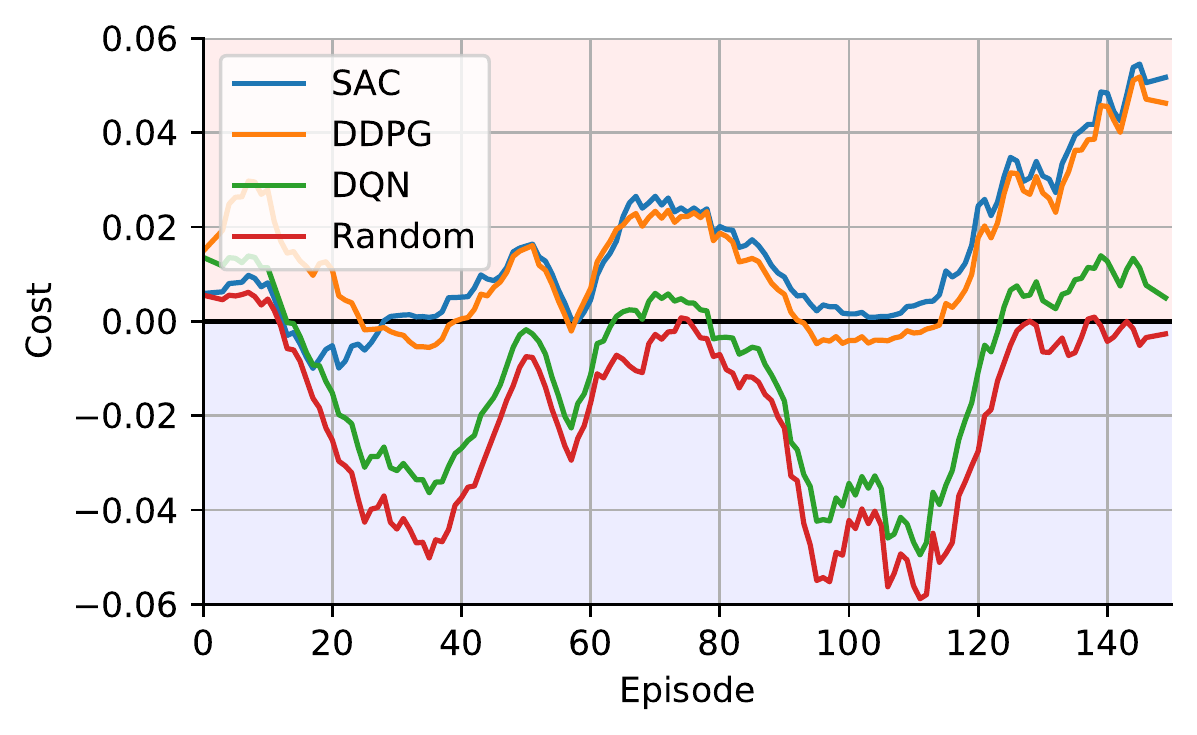}
\caption{\small The performance of different algorithms for single house.}
\label{fig:DRLperformance}
\end{figure}
Fig.~\ref{fig:DRLperformance} depicts the training procedure for a single house, comparing different local decision algorithms. In particular, we consider the proposed discretized SAC, which is based on the following stochastic policies: the actor-critic approach, discretized deep deterministic policy gradient (DDPG) a popular reinforcement learning algorithm, the DQN approach that implements standard Q-Learning procedures, and a random method as a non-learning strategy. We gradually limit the exploration capabilities of the agents in favor of the adoption of the learned policies. The variability of the scenario environment leads to experiencing learning curves with high fluctuations. Fig.~\ref{fig:DRLperformance}, is split into two blue and red parts where the blue region shows negative trading, or when the total generated energy and storage value of the battery is less than the power consumption and the energy is supplied by an external network. The red region demonstrates the positive trading where the house does not need to receive energy from the external grid. The goal is to show that DRL agents can learn the best actions and policies for energy control over a given timeframe. As expected, the SAC approach hardly copes with the definition of suitable control action policies, providing higher performances in terms of cumulative reward (cost). Similarly, DDPG suffers the temporal periodicity of the load demand, resulting in a small steep learning curve that saturates to suboptimal performances compared to SAC but generally both actor-critic methods follow a very similar trend. Conversely, the DQN and random approaches cannot have acceptable performance and consistently take correct actions according to the corresponding real-time load demands. 
\begin{figure}[h]
\centering
\includegraphics[clip, trim = 0cm 0cm 0cm 0cm,width=8cm]{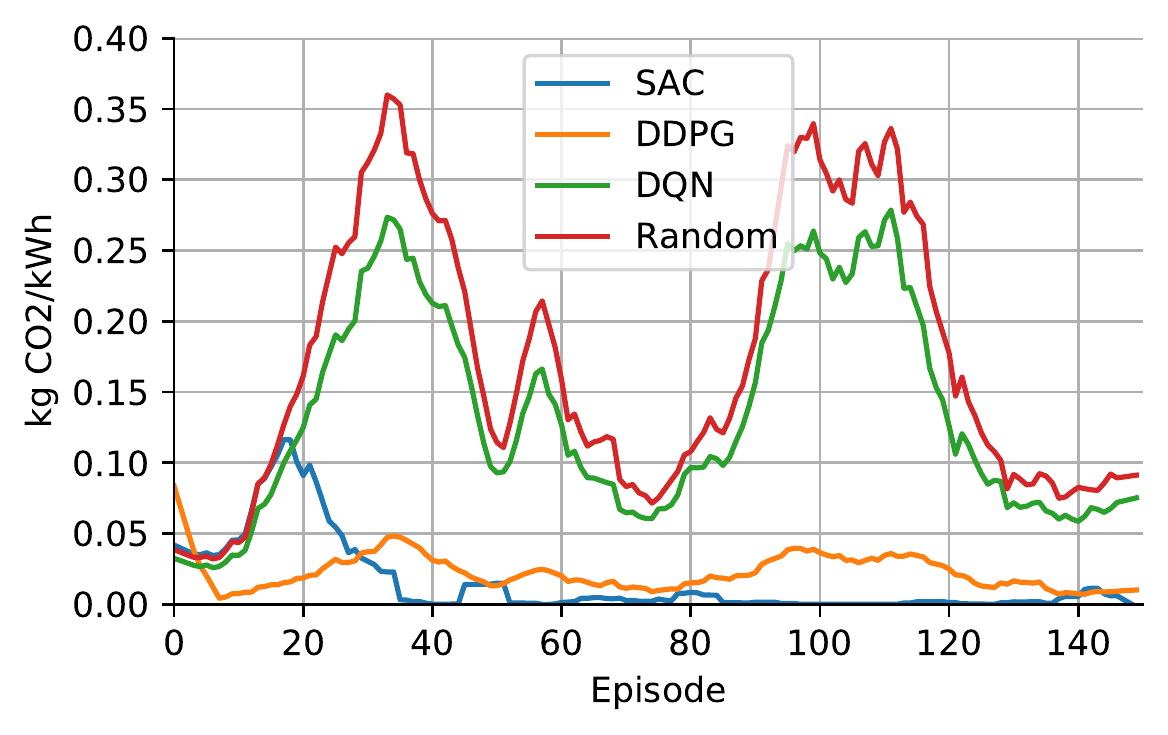}
\caption{\small The comparison of produced kg CO2/kWh for different approaches.}
\label{fig:CO2}
\end{figure}

Fig.~\ref{fig:CO2} provides an overview of the local model training procedure for the agents’ performance in terms of the amount of produced CO2, without the adoption of federate schemes. It can be noticed how DQN and random curves do not present any convergence and higher fluctuations when compared to DDPG and SAC approaches. Additionally, DQN and random curves present higher kg of CO2 emissions, suggesting a lower capability of the agents to adapt their decisions at the current load conditions. Conversely, DDPG and SAC have higher positive trading according to Fig.~\ref{fig:DRLperformance} which results in lower CO2 emissions. As shown in Fig.~\ref{fig:CO2}, the SAC method has the best performance with the agent learning the best actions of charge, discharge, and trading in a faster and robust fashion

\begin{figure}[h]
\centering
\includegraphics[clip, trim = 0cm 0cm 0cm 0cm,width=8cm]{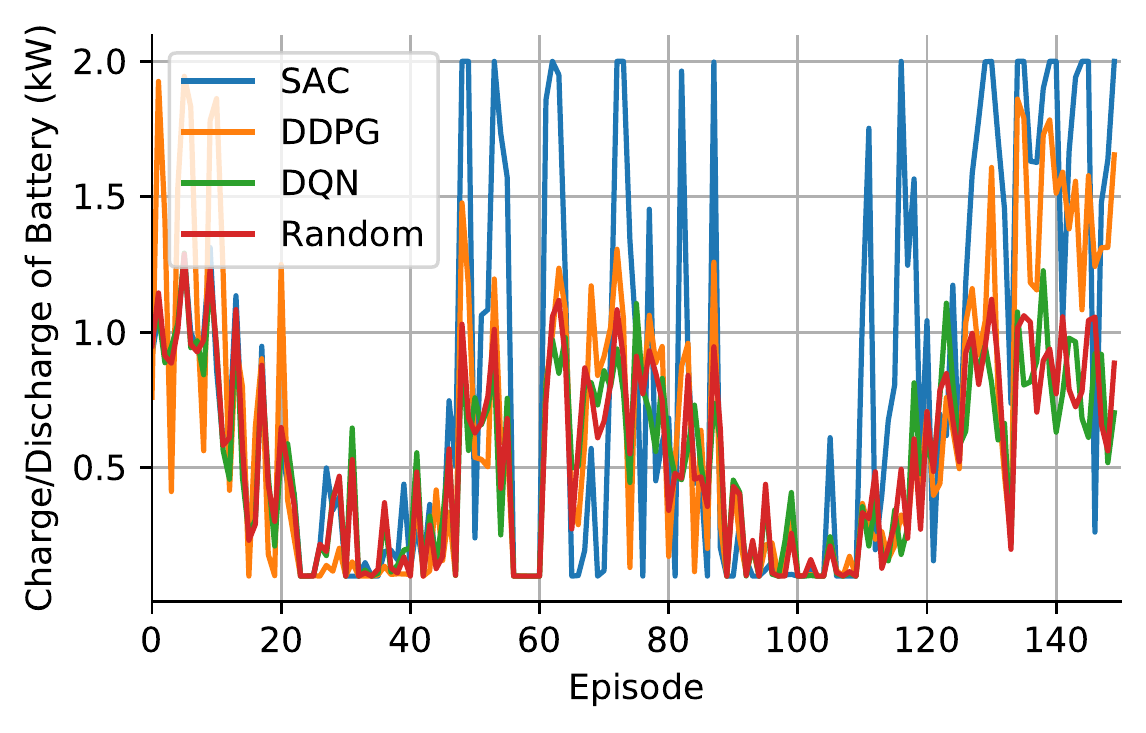}
\caption{\small The Charge/Discharge of Battery for different approaches.}
\label{fig:Battery}
\end{figure}

Fig.~\ref{fig:Battery} demonstrates a comparison between different methods in terms of battery charge/discharge in different states. It turns out that the SAC algorithm has better performance concerning the number of times that the battery is fully charged. As it could be noticed, we have defined a threshold of 10\% for battery discharging.

\begin{figure}[h]
\centering
\includegraphics[clip, trim = 0cm 0cm 0cm 0cm,width=8cm]{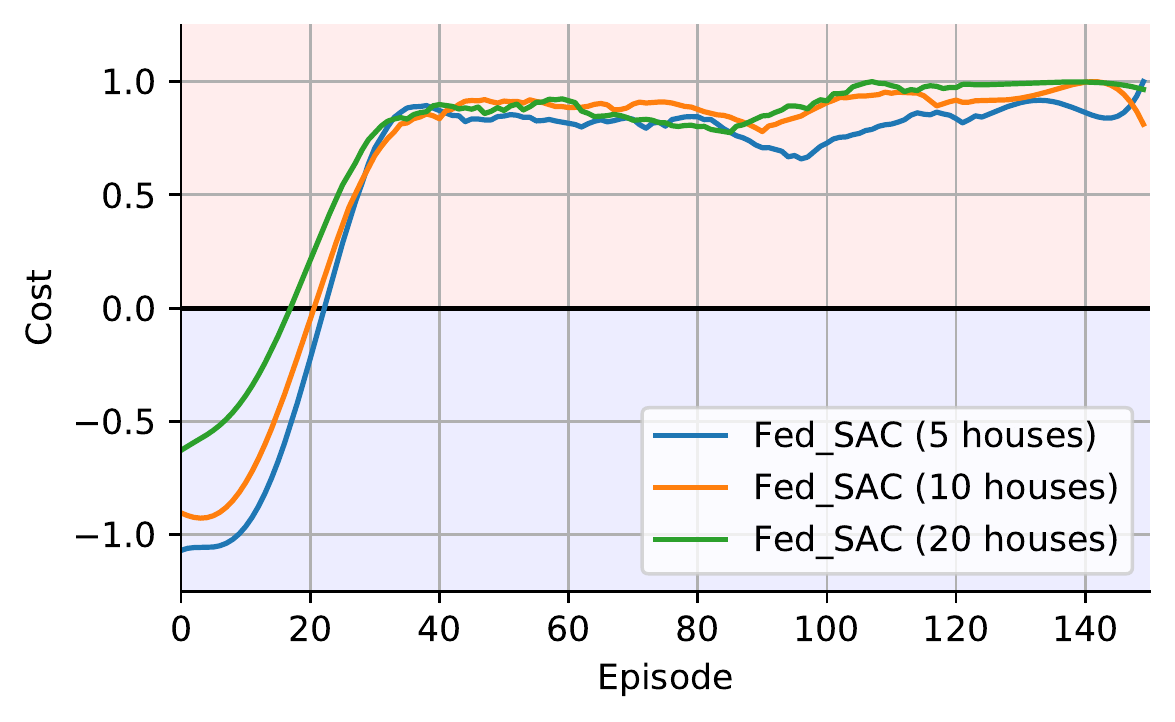}
\caption{\small The average reward of 1st house in different strategies.}
\label{fig:1st_house_performance}
\vspace{-5mm}
\end{figure}

Fig.~\ref{fig:1st_house_performance} provides a comparison of FL performances for different numbers of houses in terms of average reward.
From our experiments, it turns out that the aggregation of widely heterogeneous local models improves the capability of the global federated model to converge to a one-fits-all unified model. From the figure, we can observe how the 20 houses approach achieves better generalization of the learning policies, resulting in stable performances. Five and ten houses suffer the dynamic behavior of the underlying load conditions, presenting lower reward traces. It is worth highlighting that in terms of convergence time, in general, FDRL schemes do not necessarily provide better performances when compared to standard DRL approaches. In fact, one of the main features of FL is that it allows local DRL agents to indirectly gain knowledge on a wider state space, extending their local experience with that coming from other BEMSs deployed within the same environment. Table~\ref{tab:hyper} provides the architecture and hyperparameters of SAC method.

\begin{table}[h!]
\caption{The hyperparameters tuning in simulation}
\label{tab:hyper}
\scriptsize
\centering
\begin{tabular}{@{}lc@{}}\toprule
\textbf{Architecture} & \textbf{SAC} \\ \midrule
\textbf{Method} & Actor-Critic \\ \hdashline
\textbf{Model Type} & Multilayer perceptron\\ \hdashline
\textbf{Policy Type} & Stochastic\\ \hdashline
\textbf{Policy Evaluation} &Double Q-learning\\ \hdashline
\textbf{No. of DNNs}&6\\ \hdashline
\textbf{No. of Policy DNNs}& 1\\ \hdashline
\textbf{No. of Value DNNs}&2\\ \hdashline
\textbf{No. of Target DNNs}&3\\ \hdashline
\textbf{No. of hidden layers}& 2 \\ \hdashline
\textbf{No. of hidden units/layer}& 256 \\ \hdashline
\textbf{No. of Time Steps}& $75e3$\\ \hdashline
\textbf{Optimizer}& ADAM\\ \hdashline
\textbf{Nonlinearity}&ReLU\\ \hdashline
\textbf{Expected Entropy$(\mathcal{H})$}&-dim(Action)\\ \hdashline
\textbf{Actor Learning Rate}& 5e-4 \\ \hdashline
\textbf{Critic Learning Rate}& 5e-4\\ \hdashline
\textbf{Discount Factor}& 0.99 \\ \hdashline
\textbf{$freq$}& 2 \\ \hdashline
\bottomrule
\end{tabular}
\end{table}
\vspace{-2mm}
\section{Conclusions}
\label{sec:conclusion}

The digitization of smart grids requires more reliable management of agents and energy transactions. The smart-grid entities are moving towards decentralization, where intelligent and optimal communication between energy
prosumers and external networks can be considered a necessity. In this article, we have proposed a novel FDRL framework to coordinate the trading surpluses of energy and supply the shortages by external networks in an optimal fashion. The proposed version of SAC in BEMS trains the model and shares model hyperparameters to the federation layer in the EMS. Our simulation investigates how this federation approach is able to learn and derive the optimal policy energy control under different daytime periods, loads, and temperatures. Results show proposed algorithm has a better performance compared to other state-of-the-art DRL benchmarks while aggregation of heterogeneous local models in the form of FL improves the capability of local agents for learning optimal control actions. For future work, we will investigate the performance of the proposed approach for trading energy between smart micro-grids.

\section*{Acknowledgement} This work was partially funded by the Spanish Government (initially by MICCIN and since November 2021 by the Next Generation EU program) under Grant PCI2020-112049 and by the Electronic Components and Systems for European Leadership Joint Undertaking (JU) under grant agreement No 876868. This JU receives support from the EU`s H2020 research and innovation programme and Germany, Slovakia, Netherlands, Spain, Italy.

\end{document}